\documentclass[sigconf]{acmart}

\usepackage{amsmath,amssymb,amsfonts}
\usepackage{multirow}
\usepackage{array}
\usepackage{makecell}
\usepackage{float}
\usepackage{booktabs}
\usepackage{hyperref}
\hypersetup{hidelinks}
\usepackage{url}
\usepackage[marginal]{footmisc}
\usepackage{algorithmic}
\usepackage{graphicx}
\usepackage{subfigure}
\usepackage{textcomp}
\usepackage{xcolor}
\usepackage{colortbl}
\usepackage{tcolorbox}
\usepackage{balance}
\usepackage{threeparttable}
\usepackage{enumerate}
\usepackage{enumitem}
\AtBeginDocument{
  \providecommand\BibTeX{{
    \normalfont B\kern-0.5em{\scshape i\kern-0.25em b}\kern-0.8em\TeX}}}
\copyrightyear{2023} 
\acmYear{2023} 
\setcopyright{acmlicensed}\acmConference[EASE '23]{Proceedings of the International Conference on Evaluation and Assessment in Software Engineering}{June 14--16, 2023}{Oulu, Finland}
\acmBooktitle{Proceedings of the International Conference on Evaluation and Assessment in Software Engineering (EASE '23), June 14--16, 2023, Oulu, Finland}
\acmPrice{15.00}
\acmDOI{10.1145/3593434.3593450}
\acmISBN{979-8-4007-0044-6/23/06}
\begin{document}

\title{Code Reviewer Recommendation for Architecture Violations:\\An Exploratory Study}

\author{Ruiyin Li$^{1,2}$, Peng Liang$^{1,3*}$, Paris Avgeriou$^2$}
\affiliation{%
  \institution{$^{1}$School of Computer Science, Wuhan University, Wuhan, China}
  \institution{$^{2}$Department of Mathematics and Computing Science, University of Groningen, Groningen, The Netherlands}
  \institution{$^{3}$Hubei Luojia Laboratory, Wuhan, China}
  \institution{\{ryli\_cs, liangp\}@whu.edu.cn, p.avgeriou@rug.nl}
}

 
 

\renewcommand{\shortauthors}{Li et al.}
\renewcommand{\shorttitle}{Code Reviewer Recommendation for Architecture Violations}

\begin{abstract}
Code review is a common practice in software development and often conducted before code changes are merged into the code repository. 
A number of approaches for automatically recommending appropriate reviewers have been proposed to match such code changes to pertinent reviewers. However, such approaches are generic, i.e., they do not focus on specific types of issues during code reviews. In this paper, we propose an approach that focuses on architecture violations, one of the most critical type of issues identified during code review. 
Specifically, we aim at automating the recommendation of code reviewers, who are potentially qualified to review architecture violations, based on reviews of code changes. 
To this end, we selected three common similarity detection methods to measure the file path similarity of code commits and the semantic similarity of review comments. We conducted a series of experiments on finding the appropriate reviewers through evaluating and comparing these similarity detection methods in separate and combined ways with the baseline reviewer recommendation approach, RevFinder.
The results show that the common similarity detection methods can produce acceptable performance scores and achieve a better performance than RevFinder. The sampling techniques used in recommending code reviewers can impact the performance of reviewer recommendation approaches. We also discuss the potential implications of our findings for both researchers and practitioners.
\end{abstract}


\maketitle

\section{Introduction}\label{sec:Introduction}

Code review is widely employed in modern software development and is recognized as a valuable and effective practice at all stages of the development life cycle \cite{Bacchelli2013eoc}. Active participation of developers in code review decreases defects, improves the software quality, and facilitates knowledge sharing through rich communication among reviewers \cite{Bacchelli2013eoc, Ruangwan2018ihf}. Over the last decade, several tools have been widely used in both industry and open-source communities to make the code review process more effective, such as Phabricator\footnote{\url{https://www.phacility.com/}}, Review-Board\footnote{\url{https://www.reviewboard.org/}}, and Gerrit\footnote{\url{https://www.gerritcodereview.com/}}. Although such tools provide automated techniques to support the code review process, there is still a significant amount of human factors that can influence code review activities, such as unqualified reviewers, response delays, and overloaded review workload \cite{Chouchen2021wms, Balachandran2013rhe, Ruangwan2018ihf}.

At the heart of the human-related issues lies the process of matching code to reviewers: authors who submit new code patches to a code review system, need to invite (or the system can assign) reviewers to manually check the uploaded code fragments based on the reviewers' expertise and past experience with reviews; this may be a labor-intensive and time-consuming task, especially for large projects \cite{Cetin2021rcr}. Previous studies \cite{Bosu2016pss, Chouchen2021wms, Ruangwan2018ihf} found that effective code review requires a significant amount of effort from reviewers who thoroughly understand the submitted code. However, inappropriate code reviewers might hinder the review process, delay the incorporation of a code change into a code base, and slow down the development process. Such problems arise from misunderstanding or simply lacking knowledge of the intention or effect of code changes \cite{Dougan2019ivbt}. A proper recommendation of code reviewers can help reduce delays and speed up development by finding appropriate reviewers who are more familiar with and spend less time reviewing the submitted code fragments \cite{Thongtanunam2015wsr, Balachandran2013rhe}. 

There exist a number of code reviewer recommendation approaches in the literature (see Section \ref{sec:Reviewer Recommendation}). While these approaches can be effective, they are all generic in terms of the issues that reviewers focus. In this work, we focus on a particular type of issues: architecture violations. While architecture violations are one of the most frequently identified types of architecture issues during code review, they are not effectively covered by existing techniques and tools \cite{Li2022sae}. If code fragments with architecture violations are merged into the code base, it will increase the risk of architecture erosion \cite{Li2022SMS, Li2022sae} and gradually degrade architecture sustainability and stability \cite{Venters2018ssr}. 

The \textbf{goal} of this work is to offer an automated recommendation of code reviewers, who are potentially qualified to review architecture violations. More specifically, 
we aim at recommending potential code reviewers who have knowledge on architecture violations, through analyzing textual content of the review comments and the file paths of the reviewed code changes. Consequently, our approach is not limited to specific programming languages. This can act in a complementary way to a regular code review: a final check by reviewers who are knowledgeable in architecture violations can act as a quality gate to avoid code changes with architecture violations merged into the code base. 

Our proposed approach is novel in terms of mining semantic information in review comments from code reviewers, based on common similarity detection methods. To validate our approach, we conducted a series of experiments on 547 code review comments related to architecture violations from four Open-Source Software (OSS) projects (i.e., Nova, Neutron, Qt Base, and Qt Creator). The results show that the employed similarity detection methods can produce acceptable performance scores (i.e., values of top-\textit{k} accuracy and mean reciprocal rank metrics) and achieve a better performance than the baseline approach, RevFinder \cite{Thongtanunam2015wsr}. We managed to further explore the performance of the proposed approach on our dataset, by using fixed sampling instead of incremental sampling. The main \textbf{contributions} of our work are:
\begin{itemize}
    \item We explored the possibility of common similarity detection methods on recommending code reviewers who have awareness of architecture violations. 
    \item We conducted experiments to evaluate and compare the performance of three similarity detection methods with the baseline approach RevFinder on four OSS projects.
    \item  We shared the source code and dataset of our work \cite{Replication} to encourage further research on code reviewer recommendation for architecture issues.
\end{itemize}

The remainder of this paper is structured as follows: Section \ref{sec:Background} describes the background regarding performing code review in Gerrit, code reviewer recommendation, and architecture violations. Section \ref{sec:Research Methodology} elaborates on the research questions and study design. Section \ref{sec:Results and Discussion} presents the results of the research questions and discusses their implications. Section \ref{sec:Threats} clarifies the threats to validity and limitations of this study. Section \ref{sec:Related Work} reviews the related work and Section \ref{sec:Conclusions} concludes this study with future directions.

\section{Background}\label{sec:Background}

\subsection{Code Review Process in Gerrit}\label{sec:Code Review Practice}
\begin{figure}[t]
	\centering
	\includegraphics[width=0.65\linewidth]{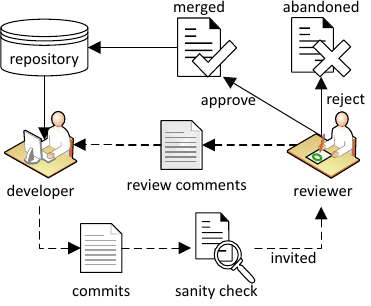}
	\caption{An overview of code review process in Gerrit}\label{F:Review Process}
\end{figure}

Code review refers to the process of inspecting source code, which is a critical activity during development and can help to improve software quality \cite{Bacchelli2013eoc}. The workflow of code review varies slightly between different platforms, e.g., the pull-request workflow in GitHub is different than code review in Gerrit. Gerrit is a commonly used platform for coordinating code review activities and facilitating traceable code reviews for git-based software development. In this work, we collected code review data from four OSS projects of two large communities OpenStack and Qt (see Section~\ref{sec:Data Collection}), both of which use Gerrit to conduct the code review process. We thus briefly elaborate on the code review process with Gerrit (see Figure \ref{F:Review Process}). 

A developer can submit new code patches or modify the original code fragments fetched from the repository through revisions; both take the form of commits. Gerrit then creates a page to record the submitted commits after conducting automated tasks, like a sanity check. Other developers of the project will be invited as reviewers to inspect the submitted commits and offer feedback (i.e., code review comments on the commits) to the developer. Such a review cycle will stop until the reviewers either approve the submitted code (status ``\textit{Merged}'') or reject it (status ``\textit{Abandoned}''). In this process, we argue that one of the code reviewers should have awareness of architecture violations and provide a review on what they are and how to fix them. Such a code review that is focused on architecture violations would be complementary to regular code reviews.

\subsection{Code Reviewer Recommendation}\label{sec:Reviewer Recommendation}
Expert recommendation is a common area in software engineering \cite{Sulun2021rst}, and code reviewer recommendation is a typical application of expert recommendation. Over the recent years, many approaches have been proposed for recommending code reviewers in the literature \cite{Thongtanunam2015wsr, Zanjani2015arp, Yu2016rrpq, Jiang2015cac, Rebai2020mcr}; these are briefly introduced below based on the categories in previous studies \cite{Cetin2021rcr, Lipcak2018als}.

\textit{Heuristic-based} approaches include problem-solving and practical methods. For example, the heuristic approaches, such as ReviewBot \cite{Balachandran2013rhe}, cHRev \cite{Zanjani2015arp}, RevFinder \cite{Thongtanunam2015wsr}, calculate heuristic scores through building expertise models to measure candidate reviewers' expertise.

\textit{Machine learning-based} approaches usually utilize data-driven machine learning techniques (e.g., Support Vector Machine (SVM) \cite{Jiang2015cac}) and genetic algorithms (e.g., Indicator-Based Evolutionary Algorithm (IBEA) \cite{Chouchen2021wms}, NSGA-II \cite{Rebai2020mcr}) to recommend reviewers. Such approaches are based on a series of features, such as patches, bug report information.

\textit{Hybrid} approaches combine different approaches (e.g., machine learning \cite{Jiang2015cac}, graph structure \cite{Yu2016rrpq}, genetic algorithm \cite{Chouchen2021wms, Rebai2020mcr}) for recommending reviewers. For example, Xia \textit{et al}. \cite{Xia2015wsr} developed a hybrid incremental approach TIE (a Text mIning and filE location-based approach) to recommend reviewers through measuring textual content (i.e., multinomial Naive Bayes, a text mining technique) and file path similarity (i.e., a VSM-based approach).

Different approaches utilize different types of artifacts to recommend code reviewers. According to the recent literature review by Çetin \textit{et al}. \cite{Cetin2021rcr}, most of the studies use pull request history (e.g., changes lines, paths of changed files and titles), and some studies also use code review history (including comments made by pull requests and reviews). Our approach can be regarded as a heuristic-based approach, and is based on both the review comments and file paths of reviewed code changes. It differs from existing approaches as it focuses specifically on architecture violations.

\subsection{Architecture Violations}\label{sec:Architecture Violation}
During software evolution, architecture erosion can degrade the stability and sustainability of system architecture due to increasing changes and accumulated architecture violations \cite{Li2022SMS, Li2021uae, Venters2018ssr}. Architecture violations are the most common and prominent type of architecture erosion symptoms; various architecture violations have been investigated in the literature \cite{Li2022SMS}. Architecture violations manifest in various ways: structural inconsistencies, violations of design decisions, violations of design principles, violations of architecture patterns, violations of API specification, etc. Previous studies on architecture violations have focused on analyzing history versions of source code. For example, Brunet \textit{et al}. \cite{Brunet2012av} carried out a longitudinal study to analyze the evolution of 19 bi-weekly versions of four OSS projects, by examining the life cycle and location of architecture violations and comparing them to the intended architecture. Maffort \textit{et al}. \cite{Maffort2016mav} proposed an approach based on defined heuristics that can rapidly raise architecture violation warnings. In contrast to the studies focusing on detecting architecture violations in source code, we aim at finding reviewers who can review architecture violations during code review, regardless of the type of architecture (e.g., micro-services, layered architecture).

\section{Research Methodology}\label{sec:Research Methodology}

\subsection{Research Questions}\label{sec:Research Questions}

\textbf{\textit{RQ1: Can common similarity detection methods be effectively used in recommending code reviewers for architecture violations?}}

This study aims at proposing an approach for the automated recommendation of code reviewers who are knowledgeable on architecture violations. To this end, we propose to use similarity measurement, as this technique is commonly used to process textual artifacts like code reviews \cite{Chouchen2021wms, Fejzer2018pbr, Ouni2016sbpr}. With this RQ, we want to investigate whether common similarity measurement techniques (i.e., Jaccard coefficient, adapted Hamming distance, cosine similarity) can indeed be useful for recommending code reviewers based on the review comments and file paths of the reviewed code changes related to architecture violations (see Section \ref{sec:Recommendation Approach}). Specifically, we plan to evaluate the performance of similarity detection methods using metrics widely adopted by the recommendation system community \cite{Cetin2021rcr}, i.e., Top-\textit{k} Accuracy and Mean Reciprocal Rank. 

\noindent\textbf{\textit{RQ2: How does the performance of the proposed similarity detection methods compare against existing code reviewer recommendation approaches?}}

With this RQ, we want to compare the similarity detection methods with an existing approach using the artifacts that we collected in our study. Specifically, there are a number of code reviewer recommendation approaches \cite{Lipcak2018als, Cetin2021rcr, Sulun2021rst}, which can be compared against the proposed approach. To be able to make the comparison, the source code of these approaches must be publicly available in order to reproduce them. 

\noindent\textbf{\textit{RQ3: Do the sampling techniques affect the performance of the proposed code reviewer recommendation approach?}}

As mentioned in a recent literature review \cite{Cetin2021rcr}, various sampling methods were used in code review recommendation. However, there are no studies that investigate whether sampling methods (see Section \ref{sec:Recommendation Approach}) can impact the performance of reviewer recommendation approaches, and which sampling techniques can achieve relatively better performance. By answering this RQ, we aim at providing empirical evidence about the influence of sampling techniques on reviewer recommendation performance.

\subsection{Data Collection}\label{sec:Data Collection}

\textbf{Projects and Code Review Comments}. 
The original dataset used in this study is from our previous work \cite{Li2022wvs}. Through a series of tasks (e.g., keywords search, manual identification and labeling of architecture violation related review comments), we collected 606 review comments on code changes and commit messages related to architecture violations. In our work, we focused on recommending reviewers who have awareness of architecture violations regarding code changes, since one of the purposes of reviewer recommendation is to help selecting reviewers for code changes. Therefore, we further extracted 547 review comments from the original dataset \cite{Li2022wvs} that are only related to code changes on architecture violations \cite{Replication}. The dataset contains the code review comments from four OSS projects, including Nova and Neutron from the OpenStack community\footnote{{\url{https://www.openstack.org/}}}, as well as Qt Base and Qt Creator from the Qt community\footnote{{\url{https://www.qt.io/}}}. As shown in Table \ref{T:Projects}, our dataset from the four OSS projects includes code review comments regarding architecture violations in eight years from June 2012 to December 2020. The review comments are related to various architecture violations (e.g., violations of design decisions, design principles, and architecture patterns), and were made by more than 200 reviewers. The items in the dataset contain review ID and patch information, including \texttt{change\_id}, \texttt{patch}, \texttt{file\_url}, \texttt{line}, and \texttt{comment}. The scripts and dataset of this work are available in \cite{Replication}.

\begin{table}[htb]
    \footnotesize
    \centering
    \renewcommand{\arraystretch}{1.3}
    \caption{Details of the selected projects used in our work}\label{T:Projects}
        \begin{threeparttable}
        \begin{tabular}{cm{22mm}<{\centering}ccc}
        \toprule
        \textbf{Project} & \textbf{Time Period} & \textbf{Files}\tnote{1} & \textbf{Comments}\tnote{2} & \textbf{Reviewers}\tnote{3} \\\hline
        Neutron    & 2013/11 - 2020/08 & 111 & 149 & 64 \\
        Nova       & 2013/01 - 2020/08 & 126 & 206 & 67 \\
        Qt Base    & 2012/12 - 2020/12 & 124 & 139 & 48 \\
        Qt Creator & 2012/06 - 2020/11 & 49  & 53  & 25 \\
        \bottomrule
        \end{tabular}
        \begin{tablenotes}
   \footnotesize
   \item[1] \textit{Files: Code change files}
   \item[2] \textit{Comments: Code review comments on architecture violations}
   \item[3] \textit{Reviewers: Code reviewers of the code change files}
        \end{tablenotes}
        \end{threeparttable}
\end{table}

\subsection{Recommendation Approach}\label{sec:Recommendation Approach}

\textbf{Problem Statement}. Since the artifacts we collected are from the OSS projects that use Gerrit as the code review tool, we take such projects as examples to formulate our approach. A software project \textit{S} contains a set of \textit{m} developers \textit{D = \{$d_1$, . . . , $d_m$\}} and \textit{n} code reviewers \textit{R = \{$r_1$, . . . , $r_n$\}}, and includes a set of \textit{j} source code files \textit{F = \{$f_1$, . . . , $f_j$\}} and a set of \textit{k} code review comments \textit{C = \{$c_1$, . . . , $c_k$\}}. In general, \textit{R} is used to represent a set of candidate code reviewers for code changes. Each reviewer $r_i$ has their own expertise on certain source code file $f_i$, and has a review comment $c_i$ on the corresponding commit. 

In such a project, each new commit (i.e., code changes that are not yet merged into the code base) could be reviewed by a number of invited (or assigned) code reviewers. Our proposal is to generate a list of recommended reviewers, in which each prospective reviewer has a matched score representing their \textit{expertise}. The higher the expertise score of a reviewer, the greater the probability for this reviewer to be recommended to review the commit. As mentioned in Section \ref{sec:Reviewer Recommendation}, our reviewer recommendation approach is based on the reviewer's \textit{expertise}, which is extracted and calculated from historical commits (e.g., review comments and file paths) and commonly used in previous studies \cite{Thongtanunam2015wsr, Chouchen2021wms, Chen2022crr, Kong2022rcr, Fejzer2018pbr}. The input of our recommendation approach is the past commit files, including file paths, review comments regarding architecture violations, and the corresponding reviewers.

\textbf{Similarity Calculation}. 
To answer RQ1 and present the \textit{expertise} of reviewers, we chose cosine similarity, Jaccard coefficient, and adapted Hamming distance to measure the similarity of file paths and the semantic similarity of review comments on architecture violations. 

In terms of the \textit{similarity of file paths}, Jaccard coefficient and adapted Hamming distance are two common methods used to measure the similarity between file paths of code changes; they are considered efficient similarity measures and widely adopted in previous studies (e.g., \cite{Chouchen2021wms, Fejzer2018pbr, Ouni2016sbpr}). Jaccard coefficient is calculated to measure similarity, as shown in Equation (\ref{equ:Jac_coefficient}):

\begin{equation}\label{equ:Jac_coefficient}
\rm Jac\_Similarity (X, Y) = \frac{\left | X \cap Y \right | }{\left | X \cup  Y \right | } 
\end{equation}

\noindent where X and Y represent two entities whose similarity needs to be measured. Here, to measure the file path similarity, X and Y represent two file paths (i.e., the sets of tokens of file paths), and the more common tokens between the file paths X and Y, the higher similarity of the two file paths.

In addition, the similarity between file paths can be also calculated by the adapted Hamming distance (i.e., \textit{similarity score = Hamming distance for the same length strings + difference in length of the two strings}). If two file paths have the same paths, then the similarity score returns 1, otherwise it returns the reciprocal score of the adapted Hamming distance of the two file paths.

In terms of the \textit{semantic similarity of code review comments}, cosine similarity and Jaccard coefficient are used to measure the semantic similarity between code review comments. The two methods are often used in previous studies (e.g., \cite{Yu2016rrpq, Rahman2017puc}) to measure the semantic similarity of textual artifacts. Cosine similarity can be utilized to determine lexical similarity between two entities represented by two vectors of words. In our case, cosine similarity is used to measure the semantic similarity of review comments regarding architecture violations, as shown in Equation (\ref{equ:cos_similarity}):

\begin{equation}\label{equ:cos_similarity}
\rm Cos\_Similarity(C_{i}, C_{j}) = \frac{v_{i}\cdot v_{j}}{\left | v_{i} \right | \left | v_{j} \right |} 
\end{equation}

\noindent where $C_{i}$ and $C_{j}$ represent two code review comments, and $v_{i}$ and $v_{j}$ denote their corresponding vectors. To generate vectors, we adopted a pre-trained Word2vec model, which was trained based on over 15 GB of textual data from Stack Overflow posts that contain a plethora of textual expressions and words in software engineering domain \cite{Efstathiou2018WE}. The higher the similarity score, the closer the two vectors that represent the two review comments.

Regarding the Jaccard coefficient, as shown in Equation (\ref{equ:Jac_coefficient}), X and Y represent two review comments in a set of tokens (i.e., tokenized words). The more common tokens between X and Y, the higher similarity score of the two review comments. Note that, before we calculated the semantic similarity by cosine similarity and Jaccard coefficient, we applied the following four pre-processing steps:
\begin{enumerate}
    \item \textit{Tokenization}. The process of tokenization is to break a stream of text into words, punctuation, and other meaningful elements called tokens.
    \item \textit{Noise removal}. Noise data usually does not contain valuable semantic information, and we therefore removed punctuation, numbers, and special characters (e.g., ``$\backslash$'', ``*'');
    \item \textit{Stop words removal}. Stop words occur commonly but do not add valuable information to differentiate different text, such as ``\textit{the}'', ``\textit{are}'', and ``\textit{is}'', which can be removed.
    \item \textit{Capitalization conversion}. We converted all the text to lower case, which can help to maintain the consistency of word form and avoid recounting the words.
\end{enumerate}

\textbf{Reviewer Recommendation}. 
For a new code change that has been commented by reviewers but has not been merged into the code base, we aim at recommending code reviewers who are potentially aware of architecture violations through measuring the similarity of the file paths and review comments of the reviewed code changes. 

We ranked the candidate code reviewers through calculating the reviewer scores using the file path similarity and the semantic similarity of historical commits (i.e., file paths and review comments of the reviewed code changes). This includes File Path similarity by Jaccard Coefficient (FP\_JC), File Path similarity by adapted Hamming Distance (FP\_HD), Review Comment semantic similarity by Cosine Similarity (RC\_CS), and Review Comment semantic similarity by Jaccard Coefficient (RC\_JC).

Given a new code change file \textit{f}, we extracted its file path \textit{$f_{new}$} and review comment \textit{$c_{new}$}, and then calculated the above-mentioned similarity scores between the current code change and each past code change, including the past file path \textit{$f_{past}$} and review comment \textit{$c_{past}$}. For example, \textit{FP\_JS($f_{new}$, $f_{past}$)} calculates the file path similarity score between \textit{$f_{new}$} and the file path \textit{$f_{past}$} of a past code change by Jaccard coefficient. Similarly, \textit{RC\_CS($c_{new}$, $c_{past}$)} calculates the semantic similarity score between \textit{$c_{new}$} and the review comment \textit{$c_{past}$} of a past review comment by cosine similarity. Then, the scores are assigned to the associated reviewers, respectively. In other words, each reviewer has four candidate similarity scores by using the four similarity detection methods. By calculating the similarity scores in separate and combined ways, a reviewer recommendation list can be generated based on the sorted reviewers along with their scores.

\begin{figure}[htb]
	\centering
	\includegraphics[width=0.9\linewidth]{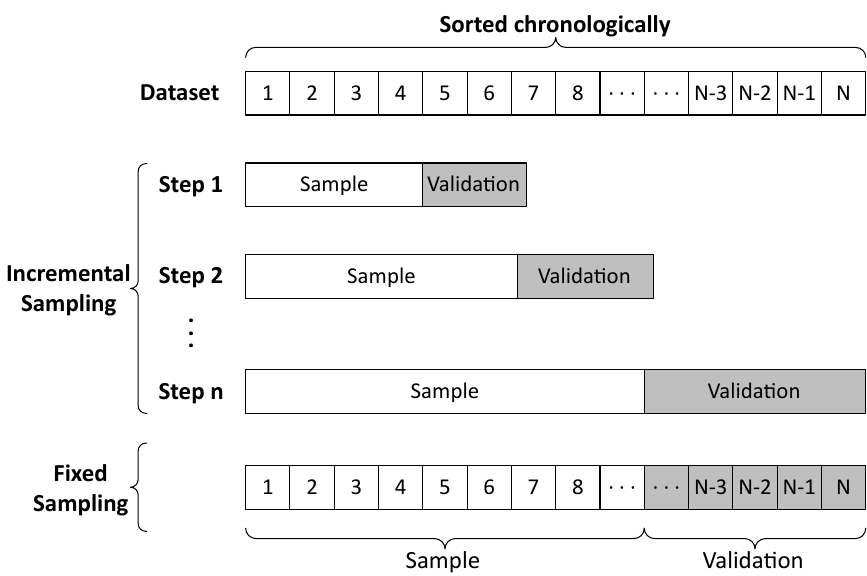}
	\caption{Overview of incremental sampling and fixed sampling}\label{F:Sampling}
\end{figure}

\textbf{Sampling and Validation}. 
Sampling refers to the sampling techniques for constructing the \textit{expertise} model, and validation denotes the process of testing the performance and effectiveness of certain sampling techniques. Unfortunately, most code reviewer recommendation studies did not provide detailed information and empirical validation on the sampling techniques for constructing their expertise models \cite{Cetin2021rcr}, and it is nontrivial to explore the \textit{effectiveness of the sampling and evaluation techniques} with the purpose of providing such empirical validation. Therefore, to answer RQ3, we planned to investigate whether and to what extent the sampling techniques can impact the performance of the proposed code reviewer recommendation approach. According to a recent literature review on code reviewer recommendation \cite{Cetin2021rcr}, incremental sampling and fixed sampling are the two most popular sampling and validation techniques (see Figure \ref{F:Sampling}), and have been commonly used in code reviewer recommendation studies \cite{Cetin2021rcr}. 

Since the code review data is temporal data, all prior studies organized their dataset chronologically \cite{Cetin2021rcr}. Thus, we also sorted our dataset in a chronological order. The incremental sampling technique takes the historical review data as the input of the expertise model through increasing the sample number in each iteration (i.e., each step in Figure \ref{F:Sampling}). The final performance of the recommendation approach is the average performance value of all the steps. In terms of the incremental sampling, we set four steps in this work, and we took 10\% of the new sample as the validation set in each step. In terms of the fixed sampling, we employed a fixed percentage of the test set by randomly sampling with 10\% in previous studies (e.g., \cite{Zubaidi2020war}) of the dataset of the four projects.

\subsection{Baseline Approach}\label{sec:Baseline}
To answer RQ2, we needed to reproduce existing expertise-based approaches, such as \cite{Yu2016rrpq, Kong2022rcr, Thongtanunam2015wsr, Chouchen2021wms}, in order to compare them against our approach. However, we were only able to do that for one approach, namely RevFinder \cite{Thongtanunam2015wsr}. The rest of the approaches had two main issues: (1) they 
require additional information that is not readily available (e.g., review workload); and (2) they do not make their source code or datasets available; some approaches did share parts of the source code, but still they can not be reproduced. This makes it difficult or even impossible to compare and evaluate our approach with these approaches.

Therefore, we reproduced one baseline approach, \textit{RevFinder}, on our collected dataset. RevFinder is a file path-based approach, which is a specific expertise-based approach and supports recommending reviewers by measuring the file path similarity of commits. Specifically, when there is a new commit, developers who have reviewed or engaged in similar revisions (i.e., with similar file paths) are likely to be recommended. Previous studies (e.g., \cite{Kong2022rcr, Chouchen2021wms, Lipcak2018als, Zanjani2015arp, Rebai2020mcr}) using various artifacts (e.g., pull-requests, historical issues) usually compared their approaches with RevFinder \cite{Thongtanunam2015wsr}, since file path is a common feature of various artifacts related to code review.

\subsection{Evaluation Metrics}\label{sec:Evaluation Metrics}
To evaluate the similarity detection methods and the baseline approach, we adopted two of the most prevalent metrics used in previous studies \cite{Cetin2021rcr}: Top-\textit{k} Accuracy and Mean Reciprocal Rank. We denote a code reviewer as \textit{r} and a code reviewer set as \textit{R}.

\textbf{Top-\textit{k} Accuracy} measures the percentage of code reviews for which an approach can properly recommend the true code reviewers within the top-\textit{k} positions in a ranked list of recommended code reviewers. In other words, this accuracy is regarding the ratio of the number of correctly recommended reviewer \textit{r} (i.e., \textit{isCorrect(r, Top-k)}) in the total number of reviewers of a ranked list of recommended reviewers. \textit{isCorrect(r, Top-k)} returns 1 if there is at least one top-\textit{k} reviewer \textit{r} who actually reviewed the code, otherwise, \textit{isCorrect(r, Top-k)} returns 0 which means a wrong recommendation. The higher the top-\textit{k} accuracy value, the better the recommendation performance. By following the previous studies in Section \ref{sec:Reviewer Recommendation}, we set the \textit{k} values of 1, 3, 5, and 10.

\begin{equation}\label{equ:Top_k}
{\rm Top\text{-}\textit{k}\ Accuracy = \frac{1}{\left | R \right |}{\sum_{r\in R}^{} isCorrect(r, Top\text{-}\textit{k})}}
\end{equation}

\textbf{Mean Reciprocal Rank (MRR)} calculates an average of reciprocal ranks of correct code reviewers in a recommendation list. Given a set of reviewers \textit{R}, MRR can be calculated by Equation (\ref{equ:MRR}). \textit{rank(r)} returns the value of the rank of the first correct reviewer in the recommendation list for reviewer \textit{r}. The value of $\frac {1}{rank(r)}$ returns 0 if there is no one who actually reviewed the code in the recommendation list. Ideally, an approach that can provide a perfect ranking should achieve an MRR value of 1. Generally, the higher the MRR value, the better the recommendation approach is.

\begin{equation}\label{equ:MRR}
{\rm MRR(R) = \frac{1}{\left | R \right |} {\sum_{r\in R}^{} \frac {1}{rank(r)}}}
\end{equation}





\begin{table*}[htb]
\scriptsize 
\centering
\setlength\tabcolsep{4.5pt}
\renewcommand{\arraystretch}{1.3}
\caption{Top\text{-}\textit{k} (1, 3, 5, 10) accuracy and MRR results of the selected similarity detection methods on four OSS projects}\label{T:Results_RQ1}
\begin{tabular}{m{30mm}<{\centering}cccccccc|cccccccc}\toprule
\textbf{Project} & \multicolumn{8}{c}{\textbf{Neutron}}& \multicolumn{8}{c}{\textbf{Nova}}\\\hline
\textbf{Similarity detection method} & \textbf{Top-1} & \textbf{MRR} & \textbf{Top-3} & \textbf{MRR} & \textbf{Top-5} & \textbf{MRR}  & \textbf{Top-10} & \textbf{MRR}  & \textbf{Top-1} & \textbf{MRR}  & \textbf{Top-3} & \textbf{MRR}  & \textbf{Top-5} & \textbf{MRR}  & \textbf{Top-10} & \textbf{MRR}  \\\hline
FP\_JC   & \cellcolor{lightgray}0.20 & 0\cellcolor{lightgray}.20 & \cellcolor{lightgray}0.33 & \cellcolor{lightgray}0.27 & \cellcolor{lightgray}0.33 & \cellcolor{lightgray}0.27 & \cellcolor{lightgray}0.33 & \cellcolor{lightgray}0.27  & \cellcolor{lightgray}0.10 & \cellcolor{lightgray}0.10 & \cellcolor{lightgray}0.14 & \cellcolor{lightgray}0.12 & \cellcolor{lightgray}0.14 & \cellcolor{lightgray}0.12 & 0.24 & 0.13 \\
FP\_HD   & \cellcolor{lightgray}0.20 & \cellcolor{lightgray}0.20 & \cellcolor{lightgray}0.33 & 0.26 & \cellcolor{lightgray}0.33 & 0.26 & \cellcolor{lightgray}0.33 & 0.26  & \cellcolor{lightgray}0.10 & \cellcolor{lightgray}0.10 & \cellcolor{lightgray}0.14 & \cellcolor{lightgray}0.12 & \cellcolor{lightgray}0.14 & \cellcolor{lightgray}0.12 & \cellcolor{lightgray}0.29 & \cellcolor{lightgray}0.14 \\
RC\_CS   & 0.07 & 0.07 & 0.20 & 0.12 & 0.30 & 0.12 & 0.27 & 0.13  & 0.00 & 0.00 & 0.00 & 0.00 & 0.05 & 0.01 & 0.19 & 0.03 \\
RC\_JC   & \cellcolor{lightgray}0.20 & \cellcolor{lightgray}0.20 & \cellcolor{lightgray}0.20 & 0.20 & 0.27 & 0.21 & \cellcolor{lightgray}0.33 & 0.22  & 0.05 & 0.05 & \cellcolor{lightgray}0.14 & 0.10 & \cellcolor{lightgray}0.14 & 0.10 & 0.24 & 0.11 \\
\multicolumn{1}{c}{\textbf{Average}} & 0.17 & 0.17 & 0.27  & 0.21 & 0.31  & 0.22 & 0.32   & 0.22 & 0.06  & 0.06 & 0.11  & 0.09 & 0.12 & 0.09 & 0.24  & 0.10 \\\hline
FP\_JC + FP\_HD   & \cellcolor{lightgray}0.27  & \cellcolor{lightgray}0.27 & \cellcolor{lightgray}0.33  & \cellcolor{lightgray}0.29 & \cellcolor{lightgray}0.33  & \cellcolor{lightgray}0.30 & \cellcolor{lightgray}0.33   & \cellcolor{lightgray}0.30 & \cellcolor{lightgray}0.10  & \cellcolor{lightgray}0.10 & \cellcolor{lightgray}0.14  & 0.12 & \cellcolor{lightgray}0.14  & 0.12 & 0.29   & \cellcolor{lightgray}0.14 \\
FP\_JC + RC\_CS   & 0.20  & 0.20 & 0.20  & 0.20 & 0.27  & 0.22 & 0.27   & 0.22 & 0.00  & 0.00 & 0.00  & 0.00 & 0.00  & 0.00 & 0.19   & 0.02 \\
FP\_JC + RC\_JC   & \cellcolor{lightgray}0.27  & \cellcolor{lightgray}0.27 & \cellcolor{lightgray}0.33  & 0.26 & \cellcolor{lightgray}0.33  & 0.29 & \cellcolor{lightgray}0.33   & 0.29 & 0.05  & 0.05 & 0.05  & 0.05 & \cellcolor{lightgray}0.19  & 0.08 & \cellcolor{lightgray}0.33   & 0.09 \\
FP\_HD + RC\_CS   & 0.13  & 0.13 & 0.13  & 0.13 & 0.27  & 0.16 & 0.27   & 0.16 & 0.00  & 0.00 & 0.00  & 0.00 & 0.05  & 0.01 & 0.19   & 0.03 \\
FP\_HD + RC\_JC   & 0.00  & 0.00 & 0.07  & 0.03 & 0.07  & 0.03 & 0.20   & 0.05 & 0.00  & 0.00 & 0.05  & 0.02 & \cellcolor{lightgray}0.19  & 0.06 & 0.24   & 0.06 \\
RC\_CS + RC\_JC   & 0.00  & 0.00 & 0.07  & 0.03 & 0.07  & 0.03 & 0.20   & 0.05 & 0.00  & 0.00 & 0.05  & 0.02 & 0.10  & 0.03 & 0.19   & 0.04 \\
\multicolumn{1}{c}{\textbf{Average}} & 0.15  & 0.15 & 0.19 & 0.16 & 0.22  & 0.17 & 0.27 & 0.18 & 0.03  & 0.03 & 0.05  & 0.04 & 0.11  & 0.05 & 0.24 & 0.06 \\\hline
FP\_HD + RC\_CS + RC\_JC   & 0.20  & 0.20 & 0.27  & 0.23 & 0.27  & 0.23 & 0.27   & 0.23 & 0.00  & 0.00 & 0.10  & 0.03 & 0.10  & 0.03 & 0.19   & 0.04 \\
FP\_JC + RC\_CS + RC\_JC   & 0.20  & 0.20 & 0.20  & 0.20 & 0.27  & 0.22 & 0.27 & 0.22 & 0.00  & 0.00 & 0.05  & 0.02 & 0.10  & 0.03 & 0.19   & 0.03 \\
FP\_JC + FP\_HD + RC\_JC   & \cellcolor{lightgray}0.27  & \cellcolor{lightgray}0.27 & \cellcolor{lightgray}0.33  & \cellcolor{lightgray}0.29 & \cellcolor{lightgray}0.33  & \cellcolor{lightgray}0.29 & \cellcolor{lightgray}0.33 & \cellcolor{lightgray}0.28 & \cellcolor{lightgray}0.05  & \cellcolor{lightgray}0.05 & \cellcolor{lightgray}0.14  & \cellcolor{lightgray}0.10 & \cellcolor{lightgray}0.19  & \cellcolor{lightgray}0.10 & \cellcolor{lightgray}0.33   & \cellcolor{lightgray}0.12 \\
FP\_JC + FP\_HD + RC\_CS   & 0.20  & 0.20 & 0.20  & 0.20 & 0.27  & 0.21 & 0.27 & 0.21 & 0.00  & 0.00 & 0.00  & 0.00 & 0.05  & 0.01 & 0.19   & 0.03 \\
\multicolumn{1}{c}{\textbf{Average}} & 0.22  & 0.22 & 0.25  & 0.23 & 0.29  & 0.24 & 0.29   & 0.24 & 0.01  & 0.01 & 0.07  & 0.04 & 0.11  & 0.04 & 0.23   & 0.06 \\\hline
FP\_JC + FP\_HD + RC\_CS + RC\_JC & 0.20  & 0.20 & 0.20  & 0.20 & 0.27  & 0.22 & 0.27   & 0.22 & 0.00  & 0.00 & 0.10  & 0.03 & 0.10  & 0.03 & 0.24   & 0.05\\\hline\hline
\textbf{Project} & \multicolumn{8}{c}{\textbf{Qt Base}}& \multicolumn{8}{c}{\textbf{Qt Creator}}\\\hline
\textbf{Similarity detection method} & \textbf{Top-1} & \textbf{MRR}  & \textbf{Top-3} & \textbf{MRR}  & \textbf{Top-5} & \textbf{MRR}  & \textbf{Top-10} & \textbf{MRR} & \textbf{Top-1} & \textbf{MRR}  & \textbf{Top-3} & \textbf{MRR}  & \textbf{Top-5} & \textbf{MRR}  & \textbf{Top-10} & \textbf{MRR}  \\\hline
FP\_JC & 0.00 & 0.00 & 0.14 & 0.05 & 0.14 & 0.05 & \cellcolor{lightgray}0.36  & 0.08 & 0.00 & 0.00 & 0.20 & 0.07 & 0.40 & 0.11 & 0.40  & 0.11 \\
FP\_HD & 0.00 & 0.00 & 0.00 & 0.00 & \cellcolor{lightgray}0.21 & 0.05 & 0.29  & 0.06 & 0.20 & 0.20 & 0.20 & 0.20 & 0.20 & 0.20 & 0.40  & 0.23 \\
RC\_CS & 0.00 & 0.00 & 0.00 & 0.00 & \cellcolor{lightgray}0.21 & 0.05 & 0.29  & 0.06 & 0.00 & 0.00 & 0.20 & 0.10 & 0.60 & 0.19 & \cellcolor{lightgray}0.80  & 0.21 \\
RC\_JC & \cellcolor{lightgray}0.07 & \cellcolor{lightgray}0.07 & \cellcolor{lightgray}0.21 & \cellcolor{lightgray}0.14 & \cellcolor{lightgray}0.21 & \cellcolor{lightgray}0.14 & 0.21  & \cellcolor{lightgray}0.14 & \cellcolor{lightgray}0.20 & \cellcolor{lightgray}0.20 & \cellcolor{lightgray}0.40 & \cellcolor{lightgray}0.27 & \cellcolor{lightgray}0.60 & \cellcolor{lightgray}0.32 & 0.60  & \cellcolor{lightgray}0.32 \\
\multicolumn{1}{c}{\textbf{Average}} & 0.02  & 0.02 & 0.09  & 0.05 & 0.19  & 0.07 & 0.29   & 0.09  & 0.10  & 0.10 & 0.25  & 0.16 & 0.45  & 0.21 & 0.55   & 0.22 \\\hline
FP\_JC + FP\_HD & 0.00 & 0.00 & 0.14 & 0.06 & 0.14 & 0.05 & 0.43  & 0.15 & 0.20 & 0.20 & 0.20 & 0.20 & \cellcolor{lightgray}0.40 & 0.25 & 0.40  & 0.25 \\
FP\_JC + RC\_CS & \cellcolor{lightgray}0.14 & \cellcolor{lightgray}0.14 & 0.14 & \cellcolor{lightgray}0.14 & 0.21 & \cellcolor{lightgray}0.16 & 0.29  & \cellcolor{lightgray}0.17 & 0.20 & 0.20 & \cellcolor{lightgray}0.40 & 0.30 & \cellcolor{lightgray}0.40 & 0.30 & \cellcolor{lightgray}0.80  & 0.36 \\
FP\_JC + RC\_JC & 0.00 & 0.00 & 0.07 & 0.04 & \cellcolor{lightgray}0.29 & 0.10 & \cellcolor{lightgray}0.43  & 0.09 & \cellcolor{lightgray}0.40 & \cellcolor{lightgray}0.40 & \cellcolor{lightgray}0.40 & \cellcolor{lightgray}0.40 & \cellcolor{lightgray}0.40 & \cellcolor{lightgray}0.40 & 0.40  & \cellcolor{lightgray}0.40 \\
FP\_HD + RC\_CS & 0.00 & 0.00 & 0.14 & 0.05 & 0.21 & 0.07 & 0.29  & 0.08 & 0.20 & 0.20 & \cellcolor{lightgray}0.40 & 0.30 & \cellcolor{lightgray}0.40 & 0.30 & \cellcolor{lightgray}0.80  & 0.35 \\
FP\_HD + RC\_JC & 0.07 & 0.07 & \cellcolor{lightgray}0.21 & 0.13 & 0.29 & 0.15 & 0.36  & 0.16 & 0.00 & 0.00 & 0.20 & 0.07 & 0.20 & 0.07 & 0.40  & 0.10 \\
RC\_CS + RC\_JC & 0.07 & 0.07 & 0.21 & 0.13 & 0.29 & 0.15 & 0.36  & 0.16 & 0.00 & 0.00 & 0.20 & 0.07 & 0.20 & 0.07 & 0.40  & 0.10 \\
\multicolumn{1}{c}{\textbf{Average}} & 0.05  & 0.05 & 0.15  & 0.09 & 0.24  & 0.11 & 0.36   & 0.14  & 0.17  & 0.17 & 0.30  & 0.22 & 0.33  & 0.23 & 0.53   & 0.26 \\\hline
FP\_HD + RC\_CS + RC\_JC   & 0.07 & 0.07 & 0.14 & 0.11 & \cellcolor{lightgray}0.29 & 0.14 & 0.29  & 0.14 & 0.20 & 0.20 & 0.40 & 0.30 & 0.40 & 0.30 & \cellcolor{lightgray}0.80  & 0.36 \\
FP\_JC + RC\_CS + RC\_JC   & 0.07 & 0.07 & 0.14 & 0.11 & 0.21 & 0.13 & 0.29  & 0.14 & \cellcolor{lightgray}0.40 & \cellcolor{lightgray}0.40 & \cellcolor{lightgray}0.40 & \cellcolor{lightgray}0.40 & \cellcolor{lightgray}0.40 & \cellcolor{lightgray}0.40 & \cellcolor{lightgray}0.80  & \cellcolor{lightgray}0.46 \\
FP\_JC + FP\_HD + RC\_JC   & 0.07 & 0.07 & 0.14 & 0.10 & 0.21 & 0.11 & \cellcolor{lightgray}0.43  & 0.14 & \cellcolor{lightgray}0.40 & \cellcolor{lightgray}0.40 & \cellcolor{lightgray}0.40 & \cellcolor{lightgray}0.40 & \cellcolor{lightgray}0.40 & \cellcolor{lightgray}0.40 & 0.40  & 0.40 \\
FP\_JC + FP\_HD + RC\_CS   & \cellcolor{lightgray}0.14 & \cellcolor{lightgray}0.14 & \cellcolor{lightgray}0.21 & \cellcolor{lightgray}0.17 & 0.21 & \cellcolor{lightgray}0.17 & 0.29  & 0\cellcolor{lightgray}.18 & \cellcolor{lightgray}0.40 & \cellcolor{lightgray}0.40 & \cellcolor{lightgray}0.40 & \cellcolor{lightgray}0.40 & \cellcolor{lightgray}0.40 & \cellcolor{lightgray}0.40 & \cellcolor{lightgray}0.80  & 0.45 \\
\multicolumn{1}{c}{\textbf{Average}} & 0.09  & 0.09 & 0.16  & 0.12 & 0.23  & 0.14 & 0.33   & 0.15  & 0.35  & 0.35 & 0.40  & 0.38 & 0.40  & 0.38 & 0.70   & 0.42 \\\hline
FP\_JC + FP\_HD + RC\_CS   + RC\_JC  & 0.07  & 0.07 & 0.21  & 0.14 & 0.21  & 0.14 & 0.29   & 0.15  & 0.40  & 0.40 & 0.40  & 0.40 & 0.40  & 0.40 & 0.80   & 0.22\\\bottomrule
\end{tabular}
\end{table*}

\section{Results and Discussion}\label{sec:Results and Discussion}

\subsection{RQ1: Effectiveness of Our Approach}\label{sec:RQ1_result}
To answer RQ1, we evaluated the performance of the similarity detection methods for recommending code reviewers. We used two similarity detection methods to measure the similarity of file paths, as well as two similarity detection methods to measure the similarity of code review comments, that is, FP\_JC, FP\_HD, RC\_CS, and RC\_JC (see Section \ref{sec:Recommendation Approach}). 
Table \ref{T:Results_RQ1} presents the performance of the similarity detection methods and their combinations.

Firstly, we evaluated the performance of the individual similarity detection methods, as shown in the four top rows per project in Table \ref{T:Results_RQ1}. The grey cells indicate the best performance metrics. The results show that the similarity detection methods yield varying results on different projects. For example, the performance of FP\_JC and FP\_HD on Neutron and Nova are better (with a higher top-\textit{k} accuracy and MRR) than on Qt Base and Qt Creator. Secondly, we evaluated the performance of the combinations of similarity detection methods on the four projects (rows 6-17 per project in Table \ref{T:Results_RQ1}). For the combinations of two similarity detection methods, the mixed similarity detection method of FP\_JC and FP\_HD achieves the best performance on Neutron and Nova projects with 0.33 accuracy at top-10 recommendation, and the mixed similarity detection method of FP\_JC and RC\_JC achieves the best top-5 (i.e., 0.33, 0.19, 0.29, and 0.40) and top-10 accuracy (i.e., 0.33, 0.33, 0.43, and 0.40) on the four projects. For the combinations of three similarity detection methods, the mixed similarity detection method of FP\_JC, FP\_HD, and RC\_JC gets the best accuracy and MRR on the four projects. In addition, we find that the performance of the combination of four similarity detection methods does not improve significantly when compared to the mixed approaches of three similarity detection methods.

Considering the \textbf{average} results of the similarity detection methods, we find that mixing three similarity detection methods can achieve a slightly higher top-\textit{k} accuracy and MRR on Neutron, Nova, and Qt Base, and a significantly better performance on Qt Creator when compared to mixing two methods. 
However, combining four similarity detection methods has no obvious performance improvement. 
In general, the results show that combining three similarity detection methods can relatively get the best performance of top-\textit{k} accuracy and MRR on the four projects.

\textbf{Discussion of RQ1}: 
The experiment results in Table \ref{T:Results_RQ1} indicate that the selected similarity detection methods can produce acceptable performance scores (MRR values between 0.06 and 0.36) on code reviewer recommendation for architecture violations on the four projects, compared to the results (MRR values between 0.14 and 0.59) of related studies on generic reviewer recommendation (e.g., \cite{Chouchen2021wms, Hu2020grc}) with more reviewer candidates (which means potentially better performance due to the larger datasets). 
Besides, we observed that the similarity detection methods can achieve varying performances on different OSS projects. One possible reason is that the effectiveness of code reviewer recommendation approach can be influenced by project characteristics (e.g., size and type of project datasets), which aligns with the findings by Chen \textit{et al}. \cite{Chen2022crr}. 
In addition, the results show that combining three similarity detection methods based on file paths and semantic information can achieve the best performance of code reviewer recommendation. We cannot observe significant improvement when combining four similarity methods. We conjecture that this is because certain similarity detection methods might affect the final performance to varying degrees on our dataset and need to be assigned with appropriate weights to their similarity values; this requires further investigation, e.g., optimizing the performance by algorithms.

The results indicate that it is still challenging to recommend code reviewers when specific issues (e.g., architecture violations) are involved, and the performance cannot be always significantly improved by combining more similarity detection methods.

\begin{center}
\fbox{\parbox{0.465\textwidth}{\textbf{Finding 1}: \textit{The performance of the similarity detection methods and their combinations can produce acceptable performance scores, and achieve varying results on different projects. Combining three similarity methods can achieve the best performance of reviewer recommendation.}}}
\end{center}

\begin{figure*}[htb]
	\centering
	\includegraphics[width=\linewidth]{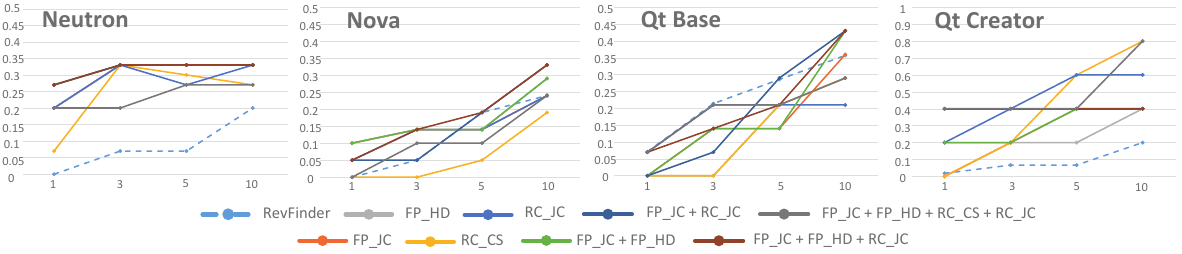}
	\caption{Performances of Top-\textit{k} accuracy of mixed similarity detection methods compared to RevFinder}\label{F:RQ2}
\end{figure*}

\begin{table}[t]
\scriptsize
\centering
\setlength\tabcolsep{6pt}
\renewcommand{\arraystretch}{1.3}
\caption{Average MRR results by the selected similarity detection methods compared with RevFinder}\label{T:MRR}
\begin{tabular}{ccccc}\toprule
\textbf{Project} & \textbf{Neutron} & \textbf{Nova} & \textbf{Qt Base} & \textbf{Qt Creator} \\\hline
RevFinder & 0.03 & 0.04 & 0.13 & 0.06 \\\hline
FP\_JC    & 0.25 & 0.12 & 0.05 & 0.07 \\
FP\_HD    & 0.24 & 0.12 & 0.03 & 0.21 \\
RC\_CS    & 0.11 & 0.01 & 0.03 & 0.13 \\
RC\_JC    & 0.21 & 0.09 & 0.12 & 0.28 \\\hline
FP\_JC + FP\_HD & 0.29 & 0.12 & 0.07 & 0.23 \\
FP\_JC + RC\_JC & 0.28 & 0.07 & 0.06 & 0.4 \\
FP\_JC + FP\_HD + RC\_JC & 0.28 & 0.09 & 0.11 & 0.4\\\hline
\multicolumn{1}{l}{FP\_JC + FP\_HD + RC\_CS + RC\_JC} & 0.21 & 0.03 & 0.13 & 0.36 \\\bottomrule
\end{tabular}
\end{table}

\subsection{RQ2: Comparison of Recommendation Approaches}\label{sec:RQ2_result}

To answer RQ2, we measured the performance of the similarity detection methods through a comparison with the baseline reviewer recommendation approach, RevFinder \cite{Thongtanunam2015wsr}. As mentioned in Section \ref{sec:Baseline}, this was the only reviewer recommendation approach that we could reproduce. Specifically, we extracted the individual methods and the best-performing combinations of the similarity detection methods from Table \ref{T:Results_RQ1}, and we compared them with RevFinder on the four OSS projects. As shown in Figure \ref{F:RQ2}, this includes the results of the four individual similarity detection methods, the combinations of two similarity detection methods (i.e., FP\_JC + FP\_HD and FP\_JC + RC\_JC), one combination of three similarity detection methods (i.e., FP\_JC + FP\_HD + RC\_JC), and one combination of four similarity detection methods (i.e., FP\_JC + FP\_HD + RC\_CS + RC\_JC). Note that Figure \ref{F:RQ2} shows different scale for Qt Creator, to observe and compare the differences of the similarity detection methods. In terms of top-\textit{k} accuracy, the four individual similarity detection methods and their combinations outperform RevFinder approximately 4 times on Neutron; the combination of FP\_JC and RC\_JC and the combination of FP\_JC, FP\_HD, and RC\_JC achieve a better top-\textit{k} accuracy than RevFinder on Nova; RevFinder achieves a relatively better top-\textit{k} accuracy than the four individual similarity detection methods on Qt Base. Nearly all the similarity detection methods and their combinations outperform RevFinder on Qt Creator.

Table \ref{T:MRR} presents the average Mean Reciprocal Rank (MRR) of the aforementioned similarity detection methods, their combinations, and RevFinder. The results show that the individual RC\_JC method can achieve a better MRR on the four projects than RevFinder. All the mixed similarity detection methods can achieve a higher MRR than RevFinder on three of the four projects (except for Nova with mixing two and three similarity detection methods). Overall, the similarity detection methods and their combinations outperform RevFinder in the majority of the cases.

\textbf{Discussion of RQ2}: 
According to the results in Figure \ref{F:RQ2} and Table \ref{T:MRR}, we find that RevFinder does not perform as good as the claimed results in the original work \cite{Thongtanunam2015wsr} when it runs on our dataset related to specific issues (i.e., architecture violations). One possible reason could be that RevFinder recommends reviewers only by comparing the file path similarity without considering the semantic similarity of related textual artifacts. Another potential reason is that the specific dataset, specifically the size and type of the dataset (review comments on architecture violations), in our work may impact the performance of reviewer recommendation. Besides, the performance of RevFinder on the four projects also partially confirms the finding of RQ1, that is, project characteristics can impact the effectiveness of reviewer recommendation approaches.

\begin{center}
\fbox{\parbox{0.465\textwidth}{\textbf{Finding 2}: \textit{The selected similarity detection methods and their combinations achieve a better performance than RevFinder in the majority of the cases}.}}
\end{center}

\begin{table*}[htb]
\scriptsize
\centering
\setlength\tabcolsep{6pt}
\renewcommand{\arraystretch}{1.4}
\caption{Top-\textit{k} accuracy by the selected similarity detection methods compared with RevFinder}\label{T:TopK_RQ3}
\begin{tabular}{ccccc|cccc||cccc|cccc}\toprule
\textbf{Project} & \multicolumn{8}{c}{\textbf{Neutron}} & \multicolumn{8}{c}{\textbf{Nova}}\\\hline
\textbf{Sampling} & \multicolumn{4}{c|}{\textbf{Fixed Sampling}} & \multicolumn{4}{c||}{\textbf{Incremental Sampling}} & \multicolumn{4}{c|}{\textbf{Fixed Sampling}} & \multicolumn{4}{c}{\textbf{Incremental Sampling}} \\\hline
\textit{\textbf{k}} & \textbf{1} & \textbf{3} & \textbf{5} & \textbf{10} & \textbf{1} & \textbf{3} & \textbf{5} & \textbf{10} & \textbf{1} & \textbf{3} & \textbf{5} & \textbf{10} & \textbf{1} & \textbf{3} & \textbf{5} & \textbf{10} \\\hline
RevFinder & 0.000 & 0.067 & 0.067 & 0.200 & 0.000 & 0.000 & 0.002 & 0.014 & 0.000 & 0.048 & 0.190 & 0.238  & 0.001 & 0.008 & 0.014 & 0.018 \\\hline
FP\_JC & 0.200 & 0.333 & 0.333 & 0.333  & 0.003 & 0.012 & 0.013 & 0.014 & 0.095 & 0.143 & 0.143 & 0.238  & 0.001 & 0.016 & 0.021 & 0.026 \\
FP\_HD & 0.200 & 0.333 & 0.333 & 0.033  & 0.003 & 0.008 & 0.013 & 0.024 & 0.095 & 0.143 & 0.143 & 0.286  & 0.001 & 0.015 & 0.018 & 0.022 \\
RC\_CS & 0.067 & 0.200 & 0.300 & 0.267  & 0.003 & 0.005 & 0.006 & 0.016 & 0.000 & 0.000 & 0.048 & 0.190  & 0.001 & 0.011 & 0.013 & 0.022 \\
RC\_JC & 0.200 & 0.200 & 0.267 & 0.333  & 0.005 & 0.018 & 0.024 & 0.024 & 0.048 & 0.143 & 0.143 & 0.238  & 0.001 & 0.010 & 0.015 & 0.016 \\\hline 
FP\_JC + FP\_HD & 0.267 & 0.333 & 0.333 & 0.333  & 0.003 & 0.008 & 0.015 & 0.016 & 0.095 & 0.143 & 0.143 & 0.286  & 0.001 & 0.015 & 0.018 & 0.023 \\
FP\_JC + RC\_JC & 0.267 & 0.333 & 0.333 & 0.333  & 0.008 & 0.012 & 0.013 & 0.014 & 0.048 & 0.048 & 0.190 & 0.333  & 0.005 & 0.016 & 0.019 & 0.024 \\\hline
FP\_JC + FP\_HD + RC\_JC & 0.267 & 0.333 & 0.333 & 0.333  & 0.008 & 0.012 & 0.012 & 0.016 & 0.048 & 0.143 & 0.190 & 0.333  & 0.003 & 0.018 & 0.019 & 0.024 \\\hline
FP\_JC + FP\_HD + RC\_CS + RC\_JC & 0.200 & 0.200 & 0.267 & 0.267  & 0.005 & 0.008 & 0.017 & 0.017 & 0.000 & 0.095 & 0.095 & 0.238 & 0.004 & 0.015 & 0.015 & 0.024\\\hline\hline
\textbf{Project} & \multicolumn{8}{c}{\textbf{Qt Base}} & \multicolumn{8}{c}{\textbf{Qt Creator}}\\\hline
\textbf{Sampling} & \multicolumn{4}{c|}{\textbf{Fixed Sampling}} & \multicolumn{4}{c||}{\textbf{Incremental Sampling}} & \multicolumn{4}{c|}{\textbf{Fixed Sampling}} & \multicolumn{4}{c}{\textbf{Incremental Sampling}} \\\hline
\textit{\textbf{k}} & \textbf{1} & \textbf{3} & \textbf{5} & \textbf{10} & \textbf{1} & \textbf{3} & \textbf{5} & \textbf{10} & \textbf{1} & \textbf{3} & \textbf{5} & \textbf{10} & \textbf{1} & \textbf{3} & \textbf{5} & \textbf{10}\\\hline
RevFinder & 0.071 & 0.214 & 0.286  & 0.357   & 0.004  & 0.022 & 0.025 & 0.038 & 0.000 & 0.200 & 0.400 & 0.400  & 0.003 & 0.006 & 0.011 & 0.033\\\hline
FP\_JC & 0.048 & 0.143 & 0.143 & 0.238  & 0.019 & 0.028 & 0.030 & 0.032 & 0.000 & 0.200 & 0.400 & 0.400  & 0.007 & 0.017 & 0.021 & 0.039 \\
FP\_HD & 0.000 & 0.000 & 0.214 & 0.286  & 0.013 & 0.013 & 0.013 & 0.026 & 0.200 & 0.200 & 0.200 & 0.400  & 0.012 & 0.012 & 0.012 & 0.029 \\
RC\_CS & 0.000 & 0.000 & 0.214 & 0.286  & 0.003 & 0.005 & 0.021 & 0.036 & 0.000 & 0.200 & 0.600 & 0.800  & 0.003 & 0.003 & 0.009 & 0.009 \\
RC\_JC & 0.071 & 0.214 & 0.214 & 0.214  & 0.010 & 0.016 & 0.023 & 0.033 & 0.200 & 0.400 & 0.600 & 0.600  & 0.000 & 0.000 & 0.000 & 0.013 \\\hline
FP\_JC + FP\_HD & 0.000 & 0.071 & 0.286 & 0.429  & 0.015 & 0.015 & 0.015 & 0.024 & 0.200 & 0.200 & 0.400 & 0.400  & 0.004 & 0.009 & 0.019 & 0.033 \\
FP\_JC + RC\_JC & 0.400 & 0.400 & 0.400 & 0.400  & 0.014 & 0.016 & 0.020 & 0.025 & 0.400 & 0.400 & 0.400 & 0.400  & 0.004 & 0.004 & 0.015 & 0.033 \\\hline
FP\_JC + FP\_HD + RC\_JC & 0.071 & 0.143 & 0.214 & 0.429  & 0.015 & 0.016 & 0.018 & 0.024 & 0.400 & 0.400 & 0.400 & 0.400  & 0.004 & 0.012 & 0.018 & 0.033 \\\hline
FP\_JC + FP\_HD + RC\_CS + RC\_JC & 0.071 & 0.214 & 0.214 & 0.286  & 0.015 & 0.017 & 0.022 & 0.040 & 0.400 & 0.400 & 0.400 & 0.800  & 0.009 & 0.009 & 0.013 & 0.013\\\bottomrule
\end{tabular}
\end{table*}

\subsection{RQ3: Comparison of Sampling Methods}\label{sec:RQ3_result}

To answer RQ3, we used the incremental sampling technique to construct the \textit{expertise} model and evaluated the performance of the selected similarity detection methods and their combinations on our dataset, as described in Section \ref{sec:Recommendation Approach}. We used the same performance metrics and the baseline approach mentioned in Section \ref{sec:Baseline} and Section \ref{sec:Evaluation Metrics}. Due to space limitations, we only present the top-\textit{k} accuracy of the best-performing similarity detection methods and their combinations in Table \ref{T:TopK_RQ3}.

According to the results in Table \ref{T:TopK_RQ3}, when using the fixed sampling technique, almost all the top-\textit{k} accuracy of the similarity detection methods and their combinations have better performance scores than when using the incremental sampling technique. For example, in terms of Neutron, the combination of FP\_JC, FP\_HD, and RC\_JC achieves top-\textit{k} accuracy of around 0.267, 0.333, 0.333, 0.333 for \textit{k} = 1, 3, 5, 10, respectively. In comparison, this combination method achieves top-\textit{k} accuracy of 0.008, 0.012, 0.012, 0.016 for \textit{k} = 1, 3, 5, 10 when using incremental sampling. Moreover, the baseline approach, RevFinder, also has such performance differences when using fixed and incremental sampling techniques. Similar observations are also valid for the MRR results of the rest of the recommendation approaches listed in Table \ref{T:TopK_RQ3}. 
In general, compared to using incremental sampling, the similarity detection methods and their combinations can achieve a better performance when using fixed sampling across all metrics.

\textbf{Discussion of RQ3}: The results of RQ3 show that all the approaches are sensitive to sampling techniques; it is clearly observed that all approaches can achieve a significantly higher recommendation performance when using the fixed sampling technique. 
One possible reason could be that the size of the dataset used to construct expertise models can influence the accuracy of reviewer recommendation. When small samples are taken as the historical review data, the first several iterations of the incremental sampling process may have relatively low performance and then decrease the final average top-\textit{k} accuracy. This conjecture corroborates the recent findings by Hu \textit{et al}. \cite{Hu2020grc} that the investigated code reviewer approaches are sensitive to training data on evaluation metrics.

\begin{center}
\fbox{\parbox{0.465\textwidth}{\textbf{Finding 3}: \textit{Sampling techniques can impact the performance of code recommendation approaches. In our work, using the fixed sampling technique to construct an expertise model can achieve a significantly better performance compared to the incremental sampling technique}.}}
\end{center}

\subsection{Implications}\label{sec:Implications}
\subsubsection{Implications for researchers}

\textbf{Establish explicit standards for research on code reviewer recommendation}. 
When a large number of submitted code changes happen, it is necessary to automate reviewer recommendation to speed up development iterations and ensure quality code reviews. In this work, we 
encountered certain issues that may hinder further research on code reviewer recommendation. For example, few existing studies chose to share their artifacts (e.g., datasets and source code). Moreover, most of the studies on reviewer recommendation are purely academic \cite{Cetin2021rcr} and lack support and validation from the industry. Besides, prior studies often use different metrics (e.g., precision, recall, and F-measure) and datasets (e.g., issues and pull-requests) in their experiments, which makes it harder to compare the performance among different approaches. Thus, it is necessary to establish explicit standards, such as promoting open science, validation in industry, and standardized metrics and datasets.

\textbf{Refine the existing approaches and focus on specific issues during code review}.
The existing approaches for code reviewer recommendation should be refined through more empirical studies. For example, developer turnover is quite common during OSS development, but is rarely considered in current studies. In addition, employing hybrid methods to recommend reviewers by combining different approaches (see Section \ref{sec:Reviewer Recommendation}) may be promising and worth exploring in the future. 
Moreover, compared to general code reviewer recommendation, recommending reviewers who have awareness of specific types of issues, such as architecture violations, is important to detect and solve these issues during code review. In this work, we conducted an exploratory study that attempts to find appropriate reviewers who have knowledge of architecture violations based on historical commits related to architecture violations. It is worth investigating other types of issues (e.g., code smells, cyclic dependencies), architectural or otherwise, in recommending reviewers with pertinent knowledge.

\subsubsection{Implications for practitioners}

\textbf{Apply and validate in industry projects}. 
Considering the characteristics of the open-source communities, the code reviewer recommendation approaches might have different performance on industrial projects; this is also pointed out in the study by Chen \textit{et al}. \cite{Chen2022crr}. Practitioners can employ reviewer recommendation approaches in their projects by taking the associated project characteristics into consideration (e.g., constructing project-specific models). More empirical validation of these approaches from industrial projects is encouraged to consolidate the findings on their performance. 
Moreover, there has been little research in code reviewer recommendation \cite{Cetin2021rcr}, and there is still a lack of industrial tools for recommending code reviewers. More collaborations between academia and industry are indispensable to devise dedicated tools (e.g., plug-ins or bots) for existing code review systems like Gerrit.

\textbf{Optimize code reviewer recommendation approaches}. 
In this work, our reviewer recommendation approach is based on the similarity of file paths and the semantic similarity of review comments. However, there are certain realistic factors that should be considered for practical software development, such as workload, availability, and developer turnover. Therefore, practitioners should pay more attention to the above-mentioned factors when optimizing the existing approaches for code reviewer recommendation. For example, they could periodically generate an updated list of candidate reviewers and add weights to the reviewers' availability.


\section{Threats to Validity}\label{sec:Threats}
In this section, we discuss the threats to the validity of this study, which may affect the results of our study.

\textbf{Construct Validity}: The main threat to construct validity in this study concerns the performance metrics (i.e., Top-\textit{k} accuracy and MRR) used in our work. 
This threat is partially mitigated in our study, as the chosen metrics are widely adopted in existing code reviewer recommendation studies \cite{Cetin2021rcr}. Besides, we shared our dataset and source code \cite{Replication} to facilitate the replication of our study and future research.

\textbf{Reliability}: The threats to reliability stem from how researchers potentially influence the study implementation. Possible threats in this study might come from the experimental settings (e.g., sampling percentage and iteration steps) and the reliability of measures (e.g., results of the similarity detection methods). As described in Section \ref{sec:Recommendation Approach}, we used a consistent and reproducible process to conduct sampling and validation on our dataset. Besides, to compare the effectiveness of our approach, we used the baseline approach (i.e., RevFinder \cite{Thongtanunam2015wsr}), which is a common baseline used in related studies. 

\textbf{External Validity}: The threats to external validity pertain to the generalizability of our results. In this study, the experiment results were produced based on the code review data of four OSS projects in Gerrit. 
Therefore, our results may not be generalized to commercial projects or open-sourced projects on other platforms (e.g., GitHub). Our future work will try to explore commercial and GitHub projects with rich code review data to better generalize the results of our approach.


\section{Related Work}\label{sec:Related Work}
Code reviewer recommendation has been gaining increasing attention in software engineering research in recent years, but there is still a lack of tools for recommending code reviewers \cite{Cetin2021rcr}. Balachandran \cite{Balachandran2013rhe} proposed automated reviewer recommendation through the ReviewBot tool, which aims at improving the review quality in an industrial context through automated static code analysis. Patanamon \textit{et al}. \cite{Thongtanunam2015wsr} proposed an expertise-based approach RevFinder based on file path similarity; their assumption is that files with similar paths have close functionality and the associated reviewers are likely to have related experience. Zanjani \textit{et al}. \cite{Zanjani2015arp} developed the cHRev approach that considers the review history including review number and review time. The cHRev approach can build an expertise model based on historical code changes and then recommend relevant peer reviewers. Yu \textit{et al}. \cite{Yu2016rrpq} provided a reviewer recommendation approach by building a social network named Comment Networks, which can capture common interests in social activities between contributors and reviewers, and then rank reviewers based on historical comments and the generated comment networks. Similarly, Kong \textit{et al}. \cite{Kong2022rcr} proposed the Camp approach based on collaboration networks along with reviewers' expertise from pull requests and file paths. 

Compared with the previous studies, our study specifically focuses on semantic information in review comments on architecture violations. We aim at recommending code reviewers who have awareness of architecture violations and can have a final check on the pending code changes (i.e., not yet being merged into the code base) that may potentially lead to architecture violations; 
this complements other reviewer recommendation approaches that can be used in combination with our approach to find pertinent (generic) code reviewers.


\section{Conclusions}\label{sec:Conclusions}
When a large number of code changes are submitted to a code review system like Gerrit, it is more efficient to find suitable code reviewers through automated reviewer recommendation compared to manually assigning reviewers. In this paper, we conducted an exploratory study to recommend qualified reviewers who have awareness of architecture violations, as a promising and feasible way to detect and prevent architecture erosion through code review.

Our study is the first attempt to explore the possibility of using similarity detection methods to recommend code reviewers on architecture violations. We evaluated the selected similarity detection methods and compared them with the baseline approach, RevFinder. The results show that the similarity detection methods and their combinations can produce acceptable performance, and the combined similarity detection methods outperform the baseline approach across most performance metrics on our dataset. Besides, we found that different sampling techniques used to build expertise models can impact the performance of code reviewer recommendation approaches, and the fixed sampling technique outperforms the incremental sampling technique on our dataset.

In the future, we plan to further optimize our reviewer recommendation approach (e.g., improve the performance through hybrid approaches discussed in Section \ref{sec:Reviewer Recommendation}) on larger datasets concerning architecture issues from diverse OSS projects and commercial systems (e.g., explore the possibility in a cross-project scenario). 

\begin{acks}
This work is funded by NSFC with No. 62172311 and the Special Fund of Hubei Luojia Laboratory.
\end{acks}

\bibliographystyle{ACM-Reference-Format}
\bibliography{ref}
\balance
\end{document}